\documentclass[12pt,preprint]{aastex}

\newcommand{\kms}{km\,s$^{-1}$\,}

\slugcomment{Accepted by ApJ}

\shorttitle{STIS Observations of the Bipolar Jet from RW Aurigae}
\shortauthors{Woitas et al.}

\begin{document}

\title{HST/STIS Observations of the Bipolar Jet from RW Aurigae:
Tracing Outflow Asymmetries Close to the Source$^1$}



\author{Jens Woitas\altaffilmark{2},
Thomas P. Ray\altaffilmark{3},
Francesca Bacciotti\altaffilmark{4},
\\
Christopher J. Davis\altaffilmark{5},
Jochen Eisl\"{o}ffel\altaffilmark{2}}

\altaffiltext{1}{Based on observations made with the
NASA/ESA {\em Hubble Space Telescope}, obtained at
the Space Telescope Science Institute,
which is operated by the Association
of Universities for Research in Astronomy,
Inc., under NASA contract NAS5-26555.}
\altaffiltext{2}{Th\"uringer Landessternwarte Tautenburg,
Sternwarte 5, D-07778 Tautenburg, Germany}
\altaffiltext{3}{School of Cosmic Physics,
Dublin Institute for Advanced Studies,
5 Merrion Square, Dublin 2, Ireland.}
\altaffiltext{4}{I.N.A.F. - Osservatorio Astrofisico di Arcetri, 
Largo E. Fermi 5, I-50125  Firenze, Italy}
\altaffiltext{5}{Joint Astronomy Centre, 660 North 
A'oh$\bar{\rm o}$k$\bar{\rm u}$ Place, 
University Park, Hilo, HI~96720} 

\begin{abstract}
We have observed the bipolar jet from RW\,Aur\,A with STIS on board the HST.
After continuum subtraction, morphological and kinematic
properties of this outflow can be traced to within 0\farcs1 from the
source in forbidden emission lines. The jet appears
well collimated, with typical FWHMs of 20 to 30 AU in the first 2$''$
and surprisingly does not show a
separate low-velocity component in contrast
to earlier observations. The systemic radial outflow velocity of the
blueshifted lobe is typically 50\% larger than
that of the redshifted one with a velocity difference of about 65\,\kms. 
Although such asymmetries
have been seen before on larger scales, our high spatial resolution 
observations suggest that they are intrinsic to the ``central engine''
rather than effects of the star's immediate environment.
Temporal variations of the bipolar jet's outflow velocities appear to occur
on timescales of a few years. They have combined to produce a 55\% increase 
in the velocity asymmetry between the two lobes over the past decade.
In the red lobe estimated mass flux $\dot{M}_j$ and  momentum flux 
$\dot{P}_j$ values are around 
one half  and one third of those for the blue lobe, respectively.
The mass outflow to mass accretion rate is 0.05, 
the former being measured at a distance of $0.''35$ from the source.
\end{abstract}



\section{Introduction} \label{intro}

Although there have been many studies of the propagation of outflows from 
young stars (e.g.\ see the reviews in {\em Protostars and Planets IV}) 
relatively little is known, from an observational perspective, about their 
generation \citep{eisletal00}. A major problem is that the source itself is 
often embedded at optical/near-infrared wavelengths 
making high resolution observations, except in the radio band, impossible. 
That said, there are a number of optically visible young stars 
associated with outflows. Although their outflows are not as striking 
as those from more embedded young stars, such systems nevertheless
represent the best window on the ``central engine'' that we have. With these 
ideas in mind, we have embarked on a Hubble Space Telescope (HST)
program to examine outflows close 
to a small number of optically visible young stars \citep{bmresc00} 
including the subject, RW\,Aur, of this {\em Letter}.  

RW\,Aur (HBC~80, HIP 23873) is a member of the Taurus-Auriga star forming
region. It is a binary system with RW\,Aur\,B  located at a position angle of
$258^{\circ}$ and a projected separation of $1\farcs50$ with respect to
RW\,Aur\,A \citep{lei1993}. Taurus-Auriga is usually assumed to have a distance
of $\approx$~140~pc (e.g.\,Wichmann et al.\,1998). In contrast the
{\em Hipparcos} distance of RW\,Aur is $70.5\pm 34.0\,\mathrm{pc}$ but,
as RW\,Aur is a close binary, and variability shifts the system's
photocentre, the latter distance has been questioned (Bertout, Robinchon, \&
Arenou 1999).
Here, we will assume 140~pc but one should keep in mind that this system
may be closer. RW\,Aur\,A is one of the optically brightest T~Tauri stars
in the sky with V = 10.1\,mag \citep{hbc1988}. The presence of strong
H$\alpha$ emission (EW[H$\alpha$] = 84 {\AA}, Herbig \& Bell 1988) and
high veiling at near-infrared wavelengths (Folha \& Emerson 1999)
categorise it as a classical T Tauri star (CTTS) with active accretion.

As is usual with CTTSs, accretion is accompanied by
outflow: \citet{hirth1994} first discovered an asymmetric bipolar jet
(HH 229) from RW\,Aur\,A using long-slit spectroscopy and found that the 
blueshifted and redshifted jets {on extended scales} differed in absolute 
systemic velocity by
a factor of two. Another unusual property of this bipolar outflow is that
the brighter lobe (in the red [SII] doublet) is redshifted. A detailed analysis
of the spectroscopic data of \citet{hirth1994} is presented in
\citet{baccetal96} who showed that the redshifted jet from RW\,Aur\,A
also has a low ionization fraction. Further observations by \citet{muneis98}
of the area surrounding this star revealed that its flow is much more
extensive than previously thought (with a total size of at least $145''$)
while \citet{doug00} examined the bipolar jet close to RW\,Aur\,A 
using adaptive optics.

Here we report on multiple observations with the Space Telescope Imaging
Spectrograph (STIS) of the bipolar jet from RW\,Aur\,A, which allow us to
spatially resolve the jet not only along the outflow direction but
transversely as well. This is the first bipolar YSO 
jet to be studied with STIS so close to its origin. Brief 
observational details are given in Section 2 and our results are presented
in Section 3.

\section{Observations and data reduction} \label{obs}

The HST/STIS observations were made on 10$^{\rm th}$ December 2000. 
Observing and data reduction procedures were virtually identical 
to those employed for DG~Tau \citep{bmresc00}. Briefly, our seven STIS spectra
of the RW\,Aur flow, taken with the G750M grating, included the 
brightest forbidden emission lines (FELs),
[OI]$\lambda\lambda$6300,6363, [NII]$\lambda\lambda$6548,6583,
[SII]$\lambda\lambda$6716,6731 along with H$\alpha$. However, highly 
blueshifted (v$_{\rm hel}\le -235$\,\kms) [OI]$\lambda$~6300 emission
was not recorded due to operating restrictions in the choice of STIS 
central wavelengths. The slit aperture was 52$\times$0.1\,arcsec$^2$, while
the spectral scale was  0.554\,\AA\,pixel$^{\rm -1}$.
The nominal spatial sampling was 0\farcs 05\,pixel$^{-1}$,
with an instrumental angular resolution of about 0\farcs1 (FWHM) 
in the red. The slit was kept parallel to the blueshifted
outflow axis (P.A.~130\degr), and offset to the southwest and northeast
of the jet axis (corresponding to left and right respectively in
the accompanying figures) in steps of 0\farcs 07, for a total coverage
in the transverse direction to the jet of about 0\farcs5. The slit was
peaked on source before offsetting. Since the position angle of the
slit is roughly perpendicular to that of RW\,Aur\,B, the presence of
this companion does not affect our observations.

We began our data reduction using the pipeline spectra employing the
most recent calibration files. After median filtering hot and dark pixels,
the continuum contribution from RW\,Aur\,A (and its associated reflection
nebula) was carefully removed. Very close to the star (i.e. less than
0\farcs1) the continuum subtraction is critical. Despite this,
the jet can be traced to at least 0\farcs1 from RW\,Aur\,A in [OI] and
[SII]. For smaller separations there is no observed forbidden line emission
above the continuum. This lack of emission is almost certainly caused by line 
quenching effects close to the star as in the case of DG~Tau \citep{bmresc00}.

Combining the spectra taken in the seven slit positions we formed images 
of the flow (``channel maps'') in eight broad velocity bins for the
lines mentioned above. The total velocity range covered is
approximately -300\,\kms to +300\,\kms which is split up into bins of
$\Delta v \approx 75$\,\kms (3 pixels). All velocities are quoted with 
respect to the mean heliocentric velocity for RW\,Aur\,A of +16\,\kms reported by 
\citet{petrov2001} and based on high velocity resolution echelle data. 
In this {\it Letter} we present interesting and surprising results
about the morphology and kinematics of the jet derived from the STIS spectra.
The spatial distribution of physical parameters such as ionisation,
hydrogen density or electron temperature, calculated from line ratios using a
diagnostic code, will be the subject of another paper (Bacciotti et al.\,
2002, in preparation).

\section{Results and Discussion} \label{res}

\begin{figure}
\epsscale{1.5}
\plotone{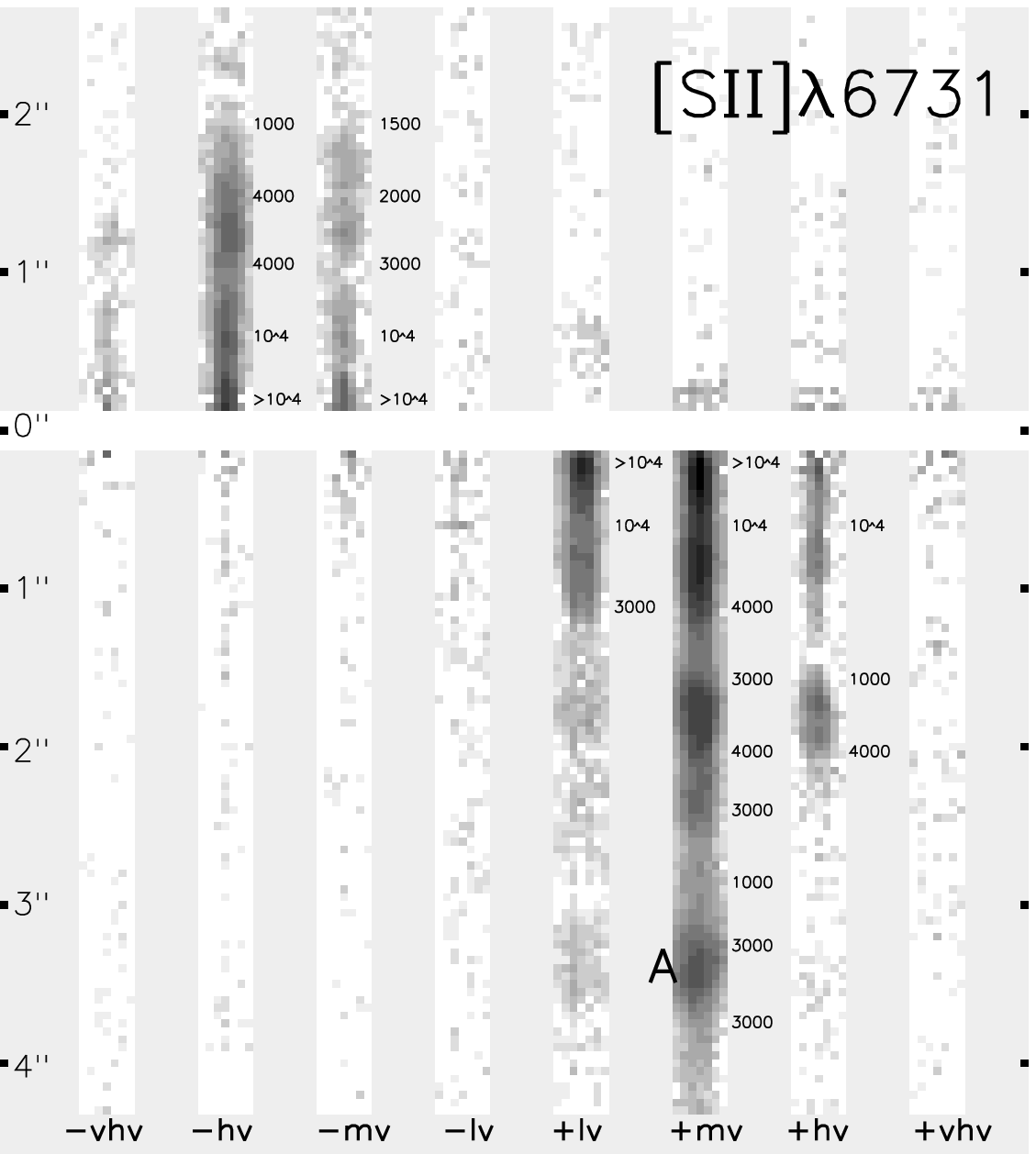}
\caption{\label{sii_lines} Images of the RW Aur jet in the
 [SII]$\lambda$6731 line reconstructed from STIS spectra
 in different radial velocity intervals. These velocities are
 heliocentric and corrected for the heliocentric velocity
 of RW~Aur (+16\,\kms, see text). The horizontal
 line marks the star's position. Each panel has a width of 0\farcs5.
 From left to right: -296\,\kms to  -221\,\kms (-vhv), \newline
 -221\,\kms to -147\,\kms (-hv), -147\,\kms to -73\,\kms (-mv), -73\,\kms to
 +1\,\kms (-lv), +1\,\kms to 75\,\kms (+lv), 75\,\kms to 150\,\kms (+mv),
 150\,\kms to 224\,\kms (+hv), 224\,\kms to 298\,\kms (+vhv). The surface
 brightness is displayed logarithmically from
$3\cdot10^{\rm -16}$ to $7\cdot10^{\rm -13}$\,erg\,s$^{\rm -1}$\,arcsec$^{\rm -2}$\,cm$^
{\rm -2}$\,\AA$^{\rm -1}$. The numeric values
 are estimates of the electron density $N_e\,[cm^{-3}]$ derived from
 the ratio of the [SII] doublet.}
\end{figure}

\begin{figure}
\epsscale{1.4}
\plotone{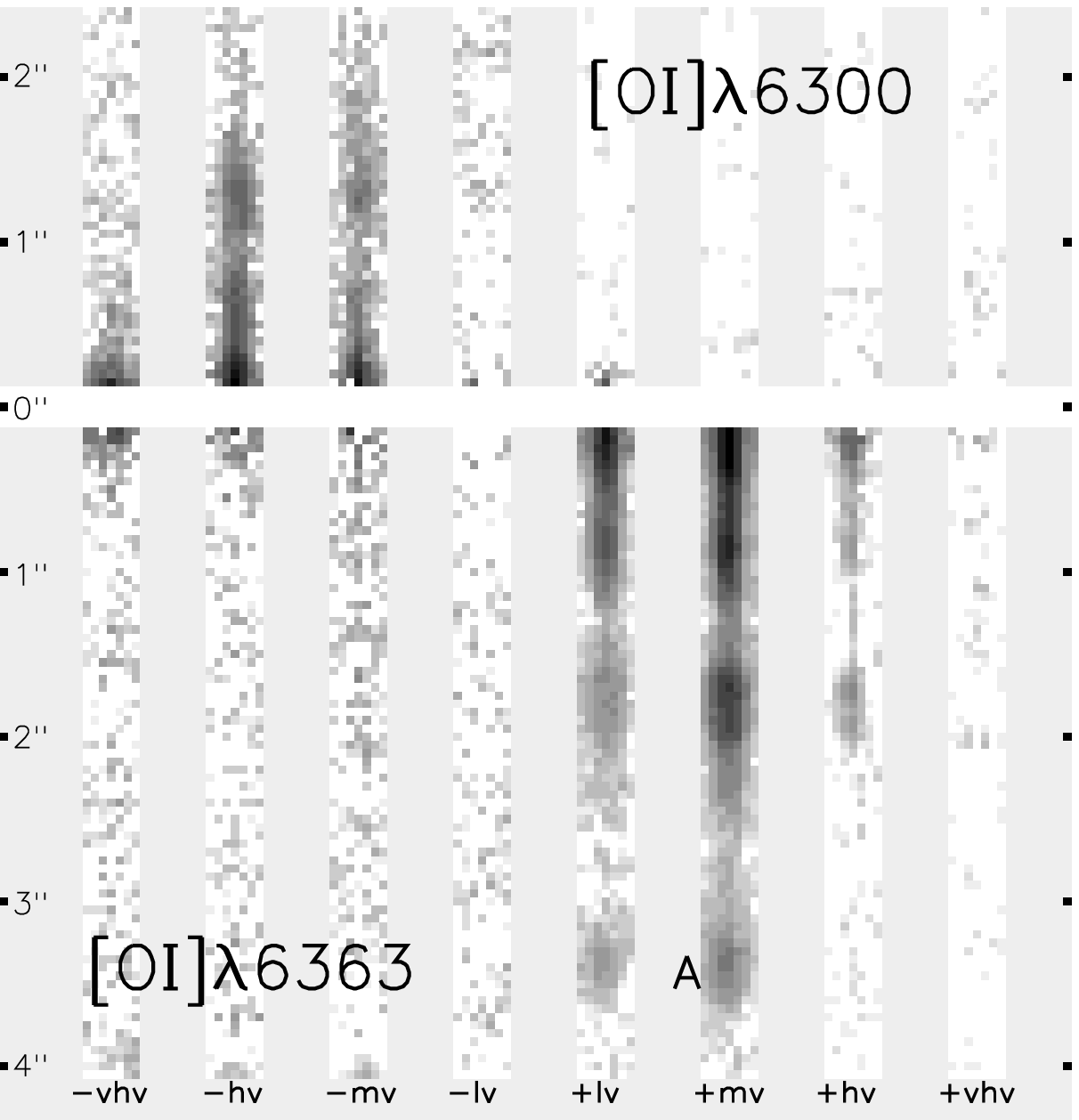}
\caption{\label{oi6300} Same as Fig.\,1, but for
 [OI]$\lambda$6300 (redshifted lobe). For the blueshifted lobe
 we use [OI]$\lambda$6363 as a substitute since the velocity profile of
 [OI]$\lambda$6300 is cut by the detector edge at $v_{\rm r}=$ -235\,\kms.
 From left to right: -318\,\kms to -240\,\kms (-vhv),
 -240\,\kms to -162\,\kms (-hv), -162\,\kms to -84\,\kms (-mv),
 -84\,\kms to -5\,\kms (-lv),
 11\,\kms to 90\,\kms (+lv), 90\,\kms to 169\,\kms (+mv),
 169\,\kms to 248\,\kms (+hv), 248\,\kms to 328\,\kms (+vhv).
 The surface brightness is displayed logarithmically from 
 $10^{\rm -15}$ to
 $1.5~\cdot~10^{\rm -12}$\,erg\,s$^{\rm -1}$\,arcsec$^{\rm -2}$\,cm$^
 {\rm -2}$\,\AA$^{\rm -1}$.}
\end{figure}

\begin{figure}
\epsscale{1.1}
\plotone{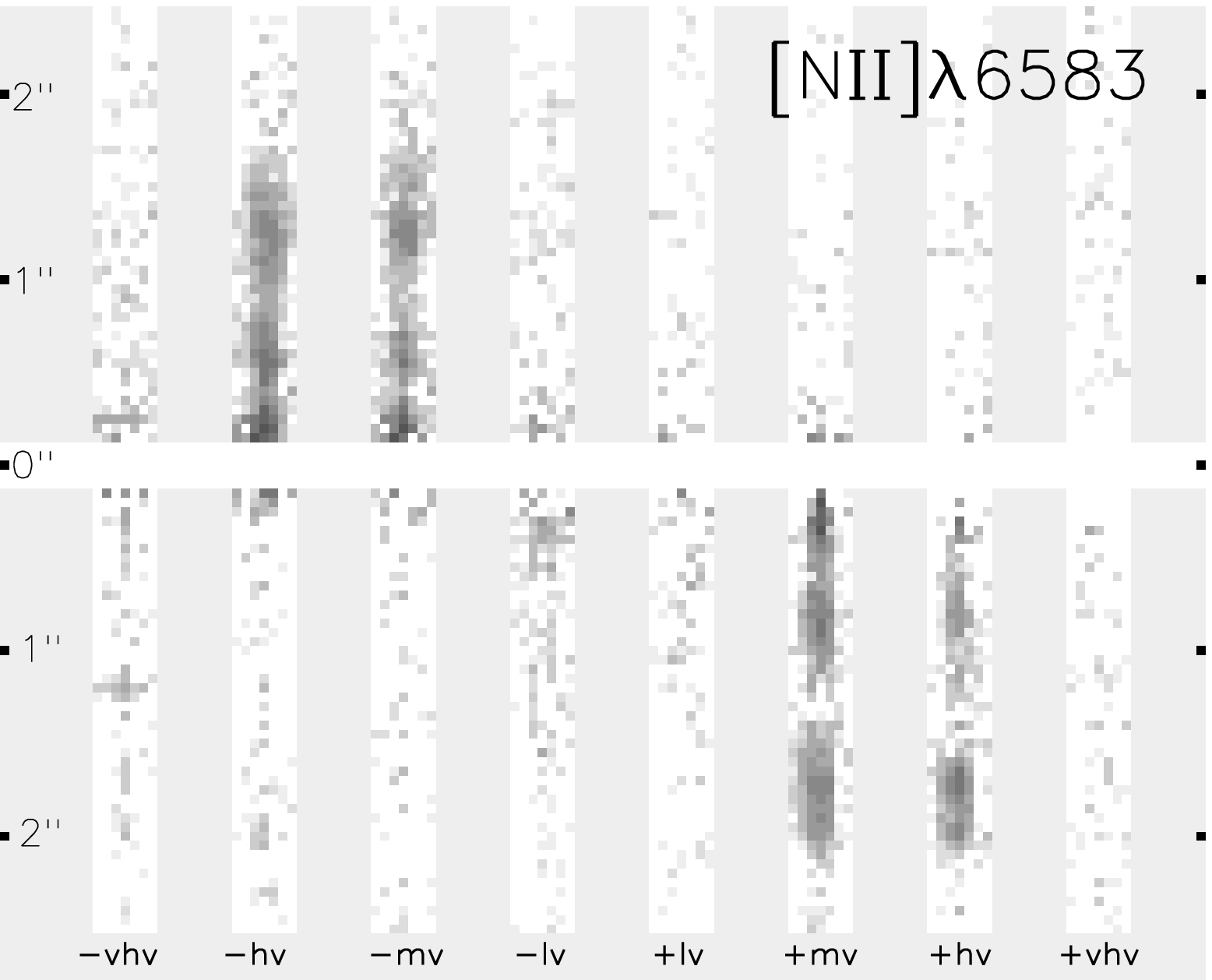}
\caption{\label{nii6583} Same as Fig.\,1, but for [NII]$\lambda$6583.
From left to right: -300\,\kms to -223\,\kms \newline (-vhv),  
-223\,\kms to -148\,\kms (-hv), -148\,\kms to -72\,\kms (-mv), -72\,\kms to 
\newline +4\,\kms (-lv), +4\,\kms to 79\,\kms (+lv), 79\,\kms to 
155\,\kms (+mv), 155\,\kms to 230\,\kms (+hv), 230\,\kms to 306\,\kms (+vhv).
The surface brightness is displayed logarithmically from
$3\cdot10^{\rm -16}$ to $3\times10^{\rm -13}$\,erg\,s$^{\rm -1}$\,arcsec$^{\rm -2}$\,cm$^{\rm -2}$\,\AA$^{\rm -1}$.}
 \end{figure}

As examples of the derived channel maps we show
[SII]$\lambda6731$ (Fig.\,\ref{sii_lines}), 
[OI]$\lambda\lambda$6300,6363 (Fig.\,\ref{oi6300}) and
[NII]$\lambda6583$ (Fig.\,\ref{nii6583}). Fig.\,\ref{sii_lines} also
contains estimates of the electron density $N_e\,[cm^{-3}]$ 
derived from the line ratio of the red [SII] doublet \citep{osterbrock1989}. 
The morphology of the jet looks
quite similar in the other lines, except for H$\alpha$ as discussed
below. Our observations provide the first set of spectra of the RW\,Aur\,A jet
with sub-arcsecond resolution. Compared to the previous study of
\citet{hirth1994}, the spatial resolution is enhanced by an order
of magnitude. \citet{doug00} provided {\em images} of the jet at 0\farcs1
angular resolution using adaptive optics and image deconvolution.
They did not detect, however, jet emission closer than about 0\farcs4
from the source. As we can trace the {\it redshifted} jet to within
approximately 15\,AU ($\approx 0\farcs1$) from the source in the FELs,
this is also an upper limit for the projected radius of any opaque disk around
RW\,Aur\,A. By implication such a disk is either seen almost edge-on
or has a radius much less than 100\,AU i.e. the usually 
assumed size of a CTTS disk (e.g. Hartmann 1998).
The latter could be the result of close encounters with RW\,Aur\,B which,
although currently separated from RW\,Aur\,A by at least 200\,AU, may
have an excentric orbit. We estimate the jet inclination angle $i$ with respect
to the line of sight from a comparison of our channel maps with the
images taken by \citet{doug00} in Dec 1997. If the jet knot at $d = 2\farcs8$
detected by \citet{doug00} in [OI] and [SII] corresponds to the
structure that we find at $d = 3\farcs4$ for intermediate velocities
in the red lobe (marked with an `A' in Figs.\,1 and 2), this indicates
a tangential velocity $v_{\mathrm{tan}}= 135$\,\kms. If we assume this 
is not a ``pattern speed'' we can combine it the with radial velocity
$v_r = 100$\,\kms for the same knot derived from our spectrum
to deduce $i = 53^{\circ}$. We do not attempt
to measure proper motions for the knots closer to the source,
because the currently observed velocity differences between knots (see below)
imply their morphology may have altered significantly (due to, for example,
knot merging) since the Dougados et al.\,(2000) observations. Given this, 
our value for $i$ should be regarded as a rough estimate. In any event even if 
the observed tangential velocity is a pattern speed it is still a {\em lower} 
limit for the actual gas tangential speed (Eisl\"offel \& Mundt 1998) and 
so it seems justified to assume $i\ge 53^{\circ}$.

	 A comparison of the forbidden line emission in the different lines 
shows that the [OI] emission appears stronger closer to the source than the 
[SII] emission. As pointed out by \citet{shang02} this is basically because 
hot dense regions contribute more to the integral
line of sight emissivity than cooler tenuous ones close to the 
star. In particular, although the electron density varies between $10^3$ and
$4\cdot 10^3\,\mathrm{cm}^{-3}$ along the jet beam, it increases
dramatically close to the source. This makes the [OI] lines brighter
than the [SII] lines at $|d|\le 0\farcs5$, since the critical density
for [OI]$\lambda\lambda$6300,6363 is $10^2$ higher than that for the [SII]
doublet. Contrary to the DG~Tau case \citep{bmresc00}, $N_e$ is higher for
moderate velocities in the red lobe. In the blue lobe, however,
the highest density is found for the high velocity bin. The jet is
rather faint in [NII] emission (Fig.\,\ref{nii6583}) in comparison
to the other FELs, and particularly with respect to
[OI]$\lambda$6300. This points towards a low ionisation fraction in
the observed portions of the jet, in agreement with results from 
previous diagnostic studies conducted at lower angular resolution
(Bacciotti et al. 1996; Dougados, Cabrit, \& Lavalley-Fouquet 2002).
Also, the [NII]/[OI] and [SII]/[OI] ratios are similar in the red and blue 
lobes, which indicates similar ionization and excitation levels \citep{be99}.

\begin{figure}
\epsscale{1.0}
\plotone{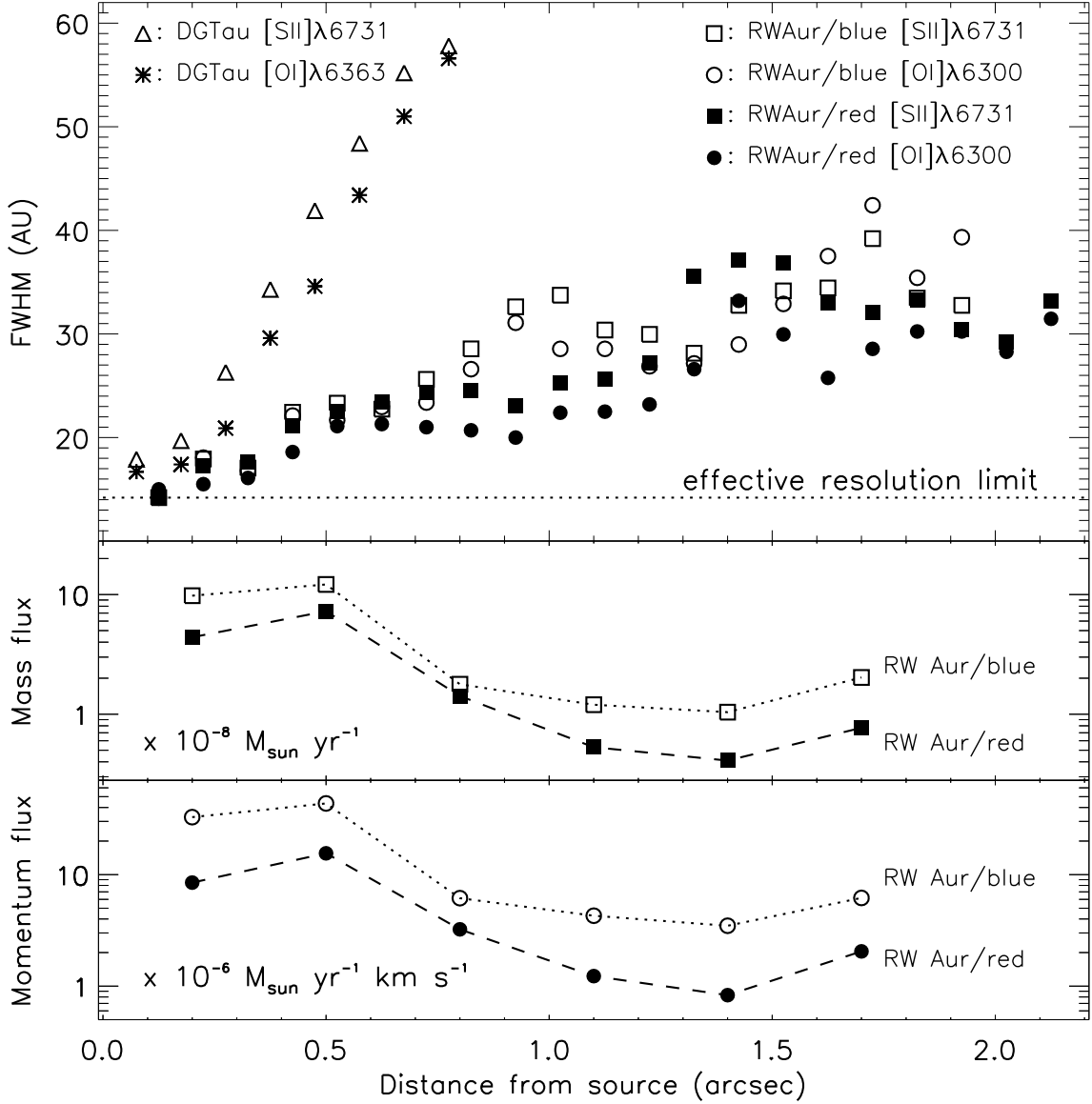}
\caption{\label{jet_width}Top Panel: Plot of FWHM width versus angular 
distance from the source for the blue and redshifted jets from RW\,Aur\,A in 
[OI]$\lambda$6300 and [SII]$\lambda$6731. All widths are measured 
binning together intermediate velocities. 
For comparison, we also show the blueshifted jet 
from DG\,Tau which is seen to have a much wider opening angle. The effective 
resolution limit is the empirical  FWHM of a reconstructed image of the 
source. Bottom Panels: Mass flux (per 10$^{-8}$ M$_{\odot}$ yr$^{-1}$) 
and momentum flux (per 10$^{-6}$ M$_{\odot}$ yr$^{-1}$ km s$^{-1}$) 
at selected positions along the blue and red lobes of the RW Aur jet.
}
\end{figure}

Both the blue and redshifted jets appear well collimated (see 
Fig.\,\ref{jet_width}, top panel). The convolved redshifted jet width 
(FWHM, measured in both [SII]$\lambda$6731 and [OI]$\lambda$6300
by binning together intermediate velocities)  
increases from 0\farcs1 close to the source to 
0\farcs2 at $d\approx $2$''$. 
In agreement with the determinations of \cite{doug00}, this 
suggests that the full redshifted jet 
opening angle (averaged over the corresponding projected distance) 
is less than 6$\degr$.
Here we also show that similar behaviour
is observed for the blueshifted jet, and 
that neither the red- or blue-shifted jets are 
well-resolved transversely at distances less than approximately 40\,AU from 
RW\,Aur\,A. This implies a jet diameter in this region of less than 15\,AU. 
For both the blue and red lobe
the values plotted in  Fig.\,\ref{jet_width} 
are marginally consistent with those obtained from the 
X-wind model (see, for example, Shang et al. 2002) and from  self-similar 
magnetised disk wind models 
\citep{garcia01}, although for the latter only with high accretion 
efficiencies. The corresponding low accretion efficiency disk wind model is 
expected to have even larger opening angles than observed 
\citep{garcia01}. Such a model might be appropriate for the  DG\,Tau jet, 
whose jet width (FWHM) values, obtained from STIS data, are 
also shown for comparison in Fig.\,\ref{jet_width} (top panel).
The figure also suggests that the electron density is typically 
higher closer to the jet axis, as the FWHM widths measured
in [OI] are almost everywhere smaller than the corresponding [SII] values.
This statement is in agreement with most models of jet acceleration.
Also, rather than having a constant opening angle, 
the jets from RW\,Aur\,A appear to expand more rapidly closer to the 
source (within 0\farcs5) than further 
away. Such behaviour was in fact suggested by groundbased observations 
over a decade ago (e.g., Mundt, Ray, \& Raga 1991) and seems to be 
supported by subsequent HST measurements (e.g., Hester, Stapelfeldt, 
\& Scowen 1998, Reipurth et al. 2000). As pointed out 
for example by \citet{reipbally01} it is not clear how to interpret such 
results: the observed jet widths may represent the extent of bow shock wings
and the true jets could be much narrower. 
We also note that farther than 0\farcs5 from the source 
there are indications for an anticorrelation between 
jet width and intensity, as already observed in HST images 
of the HH\,34 jet (Ray et al.\, 1996).

\begin{figure}
\epsscale{0.8}
\plotone{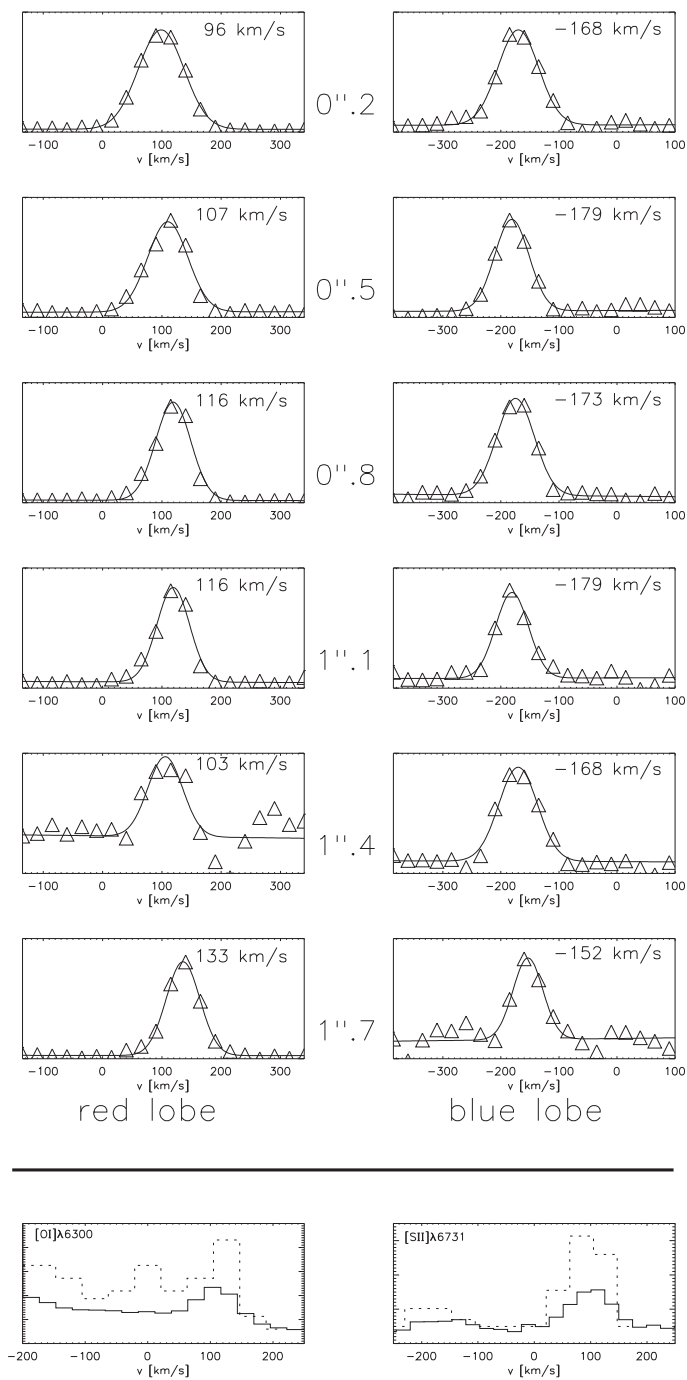}
\caption{\label{sii_vcent} Top Panel: Gaussian fits to the velocity profile of 
the red and blue lobes for the [SII]$\lambda$6731 line as derived from
 the central slit position and for different projected separations from
 the source. The numbers inside the panels are systemic peak velocities
 with uncertainties of less than 5\,\kms.
 Bottom Panel: Comparison of the [OI]$\lambda$6300 and 
[SII]$\lambda$6731 line profiles from Hartigan et al.\ (1995) (dashed) 
with the spatially averaged STIS equivalents (solid lines). For details 
see text.} 
\end{figure}

In Fig.\,\ref{sii_vcent} we present the [SII]$\lambda$6731 line profiles
at selected positions along the central slit for both the blue and red 
lobes. A Gaussian fit has been applied to find the centre. Since the
line profiles are in fact close to Gaussian and the residuals of
the fit are only a few percent, the errors in the peak velocities will
not be larger than 5\,\kms.  Over the whole range of separations covered
by our observations the velocity profiles of the lines only show one velocity
peak at $v\approx 110$~\kms in the redshifted lobe or $v\approx -170$\,\kms
in the blueshifted lobe.  In many CTTS, however,
and in addition to this high velocity component (HVC), the forbidden line
emission has another component, the so-called low-velocity component (LVC)
having typical radial velocities of $-5$ to $-20$\,\kms (e.g., Hirth, Mundt, 
\& Solf 1997) and a luminosity correlated with that of the HVC \citep{Calvet97}.
Kwan \& Tademaru (1988, 1995) have proposed that the LVC is the signature of a
poorly collimated disk wind while the HVC is a collimated jet
close to the star. In the case of RW\,Aur, LVC emission was detected by 
\citet{hartigan95} and \citet{hirth97} in the [OI]$\lambda$6300 line.
Based on observations performed in November 1993, \citet{hirth97} measured
an equivalent width ratio EW(LVC)/EW(HVC)~=~0.14. This is approximately in 
line with the correlation found by \citet{Calvet97} according to which the 
LVC flux for the [OI]$\lambda$6300 line should be $\approx 10\%$ of the 
observed HVC flux.
Although such an LVC would be easily detectable in our STIS data, it is not 
present. To further illustrate this point we have also spatially summed up our 
[OI]$\lambda$6300 and [SII]$\lambda$6731 line profile data 0\farcs5 either
side of the peak stellar continuum position. The resultant line profiles are 
shown in the upper panel of Fig.\,\ref{sii_vcent} where they are 
compared with the corresponding profiles obtained by Hartigan et al.\ (1995). 
The current lack of LVC emission is evident and indicates 
temporal variability on a timescale of a few years,
as found by \citet{Solf97} for the DG\,Tau outflow. In the context of
the Kwan \& Tademaru model this means that the RW\,Aur\,A disk wind 
(like the jet component) is variable,
perhaps as a result of magnetic instabilities (Goodson, B\"ohm, \& Winglee
1999). We also note that the absence of an LVC in RW\,Aur\,A is in 
line with recent [FeII]~1.64$\mu$m observations by \citet{davis2002}. 
The lower 
panel of Fig.\,\ref{sii_vcent} shows that the LVC was not observed by either 
Hartigan et al.\ (1995) or ourselves in [SII]$\lambda$6731. As the LVC 
often has a high electron density, this is not too surprising
giving the lower critical density of the [SII]$\lambda$6731 line in comparison
with the [OI]$\lambda$6300 line. 

In the top panel of Fig.\,\ref{sii_vcent} the radial velocity 
of the blueshifted lobe of the RW\,Aur\,A jet is seen to exceed that of the 
redshifted lobe by a factor of 
1.6 or $\approx 65$\,\kms.
This has already been recognized by \citet{hirth1994}, but only for projected
separations $\ge 300\,\mathrm{AU}$ from RW\,Aur\,A. These authors also
mentioned that radial velocity asymmetries in bipolar jets are not unusual and 
may be explained for example by different pressure gradients on both sides of 
the source. Using STIS, however we have now shown that this behaviour in the 
RW\,Aur\,A jet appears already at $d\approx 20\,\mathrm{AU}$ from the star. 
Therefore, the origin of these velocity asymmetries must be located close
to the ``central engine'' itself. Note also that asymmetries in the
excitation and opening angle of the HH~30 bipolar jet were already
evident on similar scales in WFPC2 data \citep{ray96,ber99}.

As can be seen in Fig.\,\ref{sii_vcent}, the radial outflow velocity
varies between roughly 95\,\kms and 135\,\kms in the red lobe and between
-150\,\kms and -180\,\kms in the blue lobe, in the separation range
from $d = 0\farcs2$ to $d = 1\farcs7$. This could in principle
be caused by the slit passing a ``wiggling'' jet but there is no evidence 
of wiggling in our data. The position of the jet centre, as derived from
Gaussian fits to our velocity channel maps, changes along the jet by 
less than 0\farcs02, which is much less than the slit width. In addition,
these variations show no correlation with distance from the star, suggesting
they are due to temporal variations in outflow velocity. 
As mentioned above, from a comparison of our channel maps
with images of the RW\,Aur\,A jet from Dougados et al.\,(2000) we have
estimated a jet knot proper motion of $\approx 0\farcs2\,\mathrm{yr}^{-1}$.
This implies that the observed velocity variations
occur on a timescale of 1--3 years. The knots of the RW\,Aur\,A jet
may thus be ``internal working surfaces'' produced by variations in
the outflow ejection properties as seen, for example, in the simulations of
de Gouveia Dal Pino \& Benz (1994).

The ratio $v_{\mathrm{blue}}/v_{\mathrm{red}}$
generally decreases from 1.75 at $d=0\farcs2$ to 1.14 at $d=1\farcs7$.
Under the assumptions discussed above this means that the velocity
asymmetry increased in the last decade by about 55\%. Such variations
have not been reported in the past for any bipolar YSO jet and, as far as we
are aware, this is the first time temporal evolution of bipolar jet
asymmetry has been observed. 

For both the red and the blue lobe we have
estimated the mass $\dot{M}_j$ and momentum $\dot{P}_j$ fluxes
in the jet at the distances from the source quoted in  Fig.\,\ref{sii_vcent}. 
Towards this aim, we used our determinations of velocity, 
inclination angle, electron density and jet width (FWHM), combining them with 
the estimates of the hydrogen ionization fraction 
in \cite{doug02}. The latter is reported to vary between 0.007 and 0.09 
between 0\farcs4 and 1\farcs7 in the redshifted jet, and 
we have  assumed that the ionization fraction in the blue lobe has 
the same  values. 
As already mentioned, this statement  is supported by the similarity
of the [NII]/[OI] ratio observed along the two lobes. 
We take as the dynamical diameter of the jet 
the measured FWHM, and a constant space velocity over 
the jet transverse section. The first assumption may lead to  
underestimating the fluxes, if the peripheral regions of the 
flow are too cold to be observable, but the second hypothesis should 
counter-balance the effect, since according to theoretical models 
the colder external regions are moving more slowly.
Our results are summarized in  Fig.\,\ref{jet_width}, bottom panels.
We find that in the blue lobe $\dot{M}_j$ varies from $\sim \ 1.1~10^{-7}$ 
M$_{\odot}$ yr$^{-1}$ at 0\farcs35 from the source to 
$\sim 1.2~10^{-8}$ M$_{\odot} yr^{-1}$ at 1\farcs5, while 
within the same range $\dot{P}_j$ varies from $\sim \ 3.8~10^{-5}$ 
to  $\sim 3.5~10^{-6}$ M$_{\odot}$ yr$^{-1}$ km s$^{-1}$.
In the red lobe the values for $\dot{M}_j$ and  $\dot{P}_j$  
are typically one half  and one third of those in the blue lobe, respectively.
In both lobes, the fluxes are  seen to decrease by about 
an order of magnitude in less than 2$''$ from the source. If the knots
represent mini-working surfaces, this finding 
might be justified by the fact that a considerable fraction of the
jet material is pushed sideways. Alternatively, the calculation might be 
affected from the fact that the jet is less and  less excited moving 
away from the source, and  the measured FWHM  becomes indeed substantially 
smaller than  the jet dynamical diameter. 
The mass accretion flux for this star is 1.6~10$^{-6}$ 
M$_{\odot}$ yr$^{-1}$ \citep{hartigan95}, thus 
averaging over the two lobes  one finds 
$\dot{M}_j$/$\dot{M}_{acc}$ $\sim$ 0.05 close to the source, 
which is in line with theoretical predictions of jet-launching models.
Although the origin of the blue/red  asymmetries remains unknown, 
one might ask if the observed imbalance between the momentum fluxes in 
the two lobes may cause dynamical effects on the overall system.
We have calculated that if such imbalance persisted for all the 
pre-main sequence lifetime of this star, it should have caused a 
recoil (space) velocity of 8 \kms or less for the system. 
This value, on the other hand, is probably itself 
a gross overestimate because (a) RW Aur has already accreted most of the its
mass and it may have acquired this material in a more symmetrical way, 
(b) over time linear momentum balance may be maintained. 

As a final remark, we mention that in
H$\alpha$ the appearance of the region close to the source largely
differs from the FELs. The H$\alpha$ emission can be traced back
to the stellar position in both lobes, and there are two H$\alpha$
velocity peaks on the star with the same velocities as the blue- and
redshifted lobes of the jet,
suggesting that we are tracing them back to their source.
It has been claimed (see, for example,
Hartmann 1998) that H$\alpha$ emission is not only a tracer of outflow
but of inflow as well. In particular accreted material, channelled into
funnel flows by the stellar magnetic field, may contribute to the
H$\alpha$ line profile. 
We do not see any evidence for such a contribution here. 

\section{Conclusions} \label{summ}
We have obtained spectra of the bipolar jet of RW\,Aur\,A at an unprecedented
high spatial resolution of 0\farcs1 and for the first time studied 
its morphological and kinematic properties within one arcsecond from
its origin. Both outflow lobes can be traced as close as 0\farcs1
from the source in the FELs. In H$\alpha$ the STIS spectra show
two strong maxima on the star with radial velocities coinciding
with those of the two outflow lobes. The jet appears well collimated
very close to its origin and aymmetries in the red- and blueshifted lobe
velocities arise within a region smaller than 20\,AU from the source.
This scale is almost certainly a conservative estimate as our
values for the jet inclination angle to the line of sight and
distance to RW\,Aur are lower and upper limits respectively.

In contrast to previous observations we do not find a separate low velocity
outflow component, which indicates this feature is variable on timescales
of a few years. If this component is a disk wind, then our observations 
imply such winds vary on similar timescales as the higher velocity 
outflow and may also be episodic. Variations in the high velocity components 
in the RW\,Aur\,A jet suggest that its knots may be internal working surfaces. 
This is also supported by our estimates of the mass and momentum fluxes, 
which in both lobes are observed to decrease by at least an order of 
magnitude within the first 2$''$ from the source. Close to the star 
we find an average mass flux of about 8.5 10$^{-8}$ M$_{\odot}$ yr$^{-1}$,
which is about 5\% of the mass accretion flux. The mass flux and momentum 
flux values in the red lobe are typically one half and one third of the 
values in the blue lobe, respectively.  
Finally we note that velocity
variations in the blue- and redshifted lobes imply an increase in
the radial velocity asymmetry of about 55\% over the last decade.
  




\acknowledgments
J.E.\ and J.W.\ acknowledge support by the Deutsches Zentrum f\"ur
Luft- und Raumfahrt (grant number 50 OR 0009). T.P.R. also wishes
to thank Enterprise Ireland for support (grant
SC/2000/252). The authors appreciate useful discussions with
Reinhard Mundt and the very helpful comments of an anonymous referee.

\begin{thebibliography}{}
\bibitem[Bacciotti, Hirth, \& Natta(1996)]{baccetal96} Bacciotti, F., Hirth, 
G.~A., \& Natta, A.\ 1996, \aap, 310, 309
\bibitem[Bacciotti \&  Eisl\"{o}ffel (1999)]{be99}
Bacciotti, F. \& Eisl\"{o}ffel, J., 1999, \aap, 342, 717
\bibitem[Bacciotti,  Eisl\"{o}ffel \& Ray(1999)]{ber99}
Bacciotti, F., Eisl\"{o}ffel, J., \& Ray, T.P. 1999, \aap, 350, 917
\bibitem[Bacciotti et al.(2000)]{bmresc00}
Bacciotti, F., Mundt, R., Ray, T.P., Eisl\"{o}ffel, J., Solf, J., \&
Camenzind, M. 2000, \apjl, 537, 49
\bibitem[Bertout et al.(1999)]{bertout99} Bertout, C., Robinchon, N., \&
  Arenou, F.\ 1999, \aap, 352, 574
\bibitem[Calvet(1997)]{Calvet97} Calvet, N. 1997, Proceedings of IAU
 Symposium 182, 417
\bibitem[Davis et al.(2002)]{davis2002} Davis, C.J., Whelan, E., 
Ray, T.P., \& Chrysostomou, A. 2002, in preparation
\bibitem[Dougados et al.(2000)]{doug00} Dougados, C., Cabrit, S.,
  Lavalley, C., \& M{\' e}nard, F.\ 2000, \aap, 357, L61
\bibitem[Dougados, Cabrit, \& Lavalley Fouquet(2002)]{doug02} Dougados, C., 
Cabrit, S., \& Lavalley-Fouquet, C.\ 2002, Revista Mexicana 
de Astronomia y Astrofisica Conference Series, 13, 43 
\bibitem[Eisl\"offel \& Mundt(1998)]{jochen98} Eisl\"offel, J., \& Mundt, R.
 1998, \aj, 115, 1554
\bibitem[Eisl\"offel et al.(2000)]{eisletal00} Eisl\"offel, J., Mundt, R., 
Ray, T.~P., \& Rodr\'{\i}guez, L.~F.\ 2000, Protostars and Planets IV, 815
\bibitem[Folha \& Emerson(1999)]{folha1999} Folha, D.F.M., \& Emerson, J.P. 
1999, \aap, 352, 517
\bibitem[Garcia et al.(2001)]{garcia01} 
Garcia, P.~J.~V., Cabrit, S., Ferreira, J., \& Binette, L.\ 2001, \aap, 
377, 609 
\bibitem[Goodson et al.(1999)]{goodson99} Goodson, A.~P., B\"ohm, K.~H., \&
  Winglee, R.~M. 1999, \apj, 524, 142 
\bibitem[de Gouveia Dal Pino \& Benz(1994)]{gouv94} de Gouveia Dal Pino, \&
E.M., Benz, W.\ 1994, \apj, 435, 261
\bibitem[Hartigan, Edwards, \& Ghandour(1995)]{hartigan95} 
Hartigan, P., Edwards, S., \& Ghandour, L.\ 1995, \apj, 452, 736 
\bibitem[Hartmann(1998)]{hart98} Hartmann, L., 1998, Accretion Processes in
 Star Formation (Cambridge, UK: Cambridge University Press) 
\bibitem[Herbig \& Bell(1988)]{hbc1988} Herbig, G.H., \& Bell, K.R. 1988,
  Lick Obs. Publ., 1111
\bibitem[Hester, Stapelfeldt, \& Scowen(1998)]{hester98} Hester, 
J.~J., Stapelfeldt, K.~R., \& Scowen, P.~A.\ 1998, \aj, 116, 372 
\bibitem[Hirth et al.(1994)]{hirth1994} Hirth, G.~A., Mundt, R., 
Solf, J., \& Ray, T.~P.\ 1994, \apjl, 427, L99
\bibitem[Hirth et al.(1997)]{hirth97} Hirth, G.~A., Mundt, R., \&
  Solf, J. 1997, \aaps, 126, 437
\bibitem[Kwan \& Tademaru(1988)]{kwan1988} Kwan, J., \& Tademaru, E. 1988,
 \apjl, 332, L41
\bibitem[Kwan \& Tademaru(1995)]{kwan1995} Kwan, J., \& Tademaru, E. 1995,
 \apj, 454, 382
\bibitem[Leinert et al.(1993)]{lei1993} Leinert, Ch., Zinnecker, H.,
 Weitzel, N., Christou, J., Ridgway, S.~T., Jameson, R., Haas, M., \&
 Lenzen, R. 1993, \aap, 278, 129
\bibitem[Mundt \& Eisl\"offel(1998)]{muneis98} Mundt, R., \&
 Eisl\"offel, J.\ 1998, \aj, 116, 860
\bibitem[Mundt, Ray, \& Raga(1991)]{mundt91} Mundt, R., Ray, 
T.~P., \& Raga, A.~C.\ 1991, \aap, 252, 740 
\bibitem[Osterbrock(1989)]{osterbrock1989} Osterbrock, D.E. 1989,
 Astrophysics of Gaseous Nebulae and Active Galactic Nuclei,
 University Science Books, Mill Valley, CA
\bibitem[Petrov et al.(2001)]{petrov2001} Petrov, P.~P., Gahm, G.~F., 
 Gameiro, J.~F., Duemmler, R., Ilyin, I.~V., Laakonen, T., Lago, M.~T.~V.~T, \&
 Tuominen, I. 2001, \aap, 369, 993 
\bibitem[Ray et al.(1996)]{ray96} Ray, T.~P., Mundt, R., Dyson, J.~E.,
 Falle, S.~A.~E.~G., \& Raga, A.~C. 1996, \apjl, 468, L103
\bibitem[Reipurth \& Bally(2001)]{reipbally01} Reipurth, B.~\& 
Bally, J.\ 2001, \araa, 39, 403 
\bibitem[Reipurth et al.(2000)]{reipurth00} Reipurth, B., 
Heathcote, S., Yu, K., Bally, J., \& Rodr\'{\i}guez, L.~F.\ 
2000, \apj, 534, 317 
\bibitem[Shang et al.(2002)]{shang02} 
Shang, H., Glassgold, A.~E., Shu, F.~H., \& Lizano, S.\ 2002, \apj, 564, 853 
\bibitem[Solf(1997)]{Solf97} Solf, J. 1997, Proceedings of IAU Symposium
 182, 63
\bibitem[Wichmann et al.(1998)]{wichetal98} Wichmann, R., Bastian, U.,
 Krautter, J., Jankovics, I., \& Ruci\'{n}sky, S.~M. 1998, MNRAS, 301, L39
\end{thebibliography}
\end{document}